\documentclass[twocolumn, times, tighten]{aastex63}
\usepackage{graphicx,multirow,hyperref,url,color}
\hypersetup{colorlinks=true, urlcolor=blue, citecolor=blue}

\newcommand{\kev}{\rm \,keV}
\newcommand{\lx}{L_{\rm X}}
\newcommand{\lr}{L_{\rm R}}
\newcommand{\ergs}{{\rm \,erg\,s^{-1}}}
\newcommand{\ergscm}{{\rm \,erg\,s^{-1}\,cm^{-2}}}

\newcommand{\mbh}{M_{\rm BH}}
\newcommand{\msun}{{\rm M}_{\sun}}
\newcommand{\ledd}{L_{{\rm Edd}}}

\newcommand{\xmm}{{\textit{XMM--Newton}}}
\newcommand{\rxte}{{\textit{RXTE}}}
\newcommand{\nustar}{{\textit{NuSTAR}}}
\newcommand{\swift}{{\textit{Swift}}}

\begin{document}

%\received{...}
%\revised{...}
%\accepted{...}
%% Command to document which AAS Journal the manuscript was submitted to.
%% Adds "Submitted to " the argument.
%\submitjournal{ApJ}

\title{Radio/X-ray Correlation in the Mini-Outburts of Black Hole X-ray Transient GRS 1739--278}
\shorttitle{Flat Radio/X-ray Correlation}

\author[0000-0001-9969-2091]{Fu-Guo Xie}
\affil{Key Laboratory for Research in Galaxies and Cosmology, Shanghai Astronomical Observatory, Chinese Academy of Sciences, \\
80 Nandan Road, Shanghai 200030, China; \href{mailto:fgxie@shao.ac.cn}{fgxie@shao.ac.cn}, \href{mailto:zyan@shao.ac.cn}{zyan@shao.ac.cn}}

\author[0000-0002-5385-9586]{Zhen Yan}
\affil{Key Laboratory for Research in Galaxies and Cosmology, Shanghai Astronomical Observatory, Chinese Academy of Sciences, \\
80 Nandan Road, Shanghai 200030, China; \href{mailto:fgxie@shao.ac.cn}{fgxie@shao.ac.cn}, \href{mailto:zyan@shao.ac.cn}{zyan@shao.ac.cn}}

\author{Zhongzu Wu}
\affil{College of Science, Guizhou University, Guiyang 550025, Guiyang, China; \href{mailto:zzwu08@gmail.com}{zzwu08@gmail.com}}

\correspondingauthor{Zhongzu Wu, Zhen Yan, Fu-Guo Xie}

\shortauthors{Xie, Yan \& Wu}

\begin{abstract}
We present quasi-simultaneous radio and X-ray observations of the black hole X-ray binary GRS 1739--278 of its 2015-2016 mini-outbursts, i.e. between 2015 June 10 and 2016 October 31, with the X-ray-to-radio time interval being less than one day. The monitor campaign was run by \swift\, in the X-rays and by JVLA in the radio (at both 5 GHz and 8 GHz). We find that the brightest radio emission is actually achieved during the soft sate, and the spectrum is marginally optically-thick with the spectral index $\alpha\approx -0.28\pm0.17$ (flux $F_\nu\propto \nu^\alpha$). For the radio emission in the hard state, we find a large diversity in the spectral index, i.e. a majority of radio spectra are optically-thick with $-0.5 \la \alpha \la 0.5$, while a few are optically-thin, with $\alpha$ being lower than $-1$ in certain cases. We then investigate the correlation between the luminosities in radio (monochromatic at 5 GHz, $\lr$) and 1-10 keV X-rays ($\lx$) during the hard state. We find that for more than two orders of magnitude variation in the X-ray luminosity, this source exhibits a flat correlation with $p\approx 0.16$ (in the form of $\lr\propto\lx^p$), i.e. it belongs to the ``outlier'' (to the standard correlation with $p\approx 0.6$) category that may follow a hybrid correlation. Both the slope and the corresponding luminosity range agree well with those in H1743--322, the prototype of the hybrid correlation. Theoretical implications of our results are discussed.
\end{abstract}


\section{Introduction}
\label{sec:intro}

A majority of stellar-mass black hole (BH) X-ray binaries (BHBs) are transients. After a long period of quiescence, they occasionally undergo outbursts. According to the spectral and timing properties, two distinctive states are identified \citep{Zdziarski04, RM06, Done07, Belloni10}. One is the soft state, in which the X-ray spectrum is dominated by thermal component from cold accretion disc \citep{SS73}, and is supplemented by a weak power-law tail with photon index $\Gamma > 2.1$ (X-ray spectrum defined as $F_\nu \propto \nu^{1-\Gamma}$) hard X-rays. The other is the hard state, in which the X-ray spectrum is dominated by emission from the Compton scattering within hot accretion flows, while the thermal emission becomes much weaker. Observationally, the Comptonized emission is shown as a $\Gamma \approx 1.4$-1.8 power-law emission which has an exponential high-energy cutoff at around $\sim$100 \kev\, (e.g., \citealt*{Zdziarski98}). Additionally when the thermal and the non-thermal emissions are of comparable significance, it is defined as the intermediate state, which is further divided into hard intermediate and soft intermediate states. 

BHBs in their hard state are ubiquitous with compact self-absorbed radio emission that originates from highly-collimated relativistic jet (for reviews, see \citealt*{Corbel04, Fender04, Fender09, Fender14}). This so-called ``continuous jet'' or ``steady jet'' has optically-thick radio spectrum, with $\alpha \sim 0$ (defined as flux $F_\nu\propto \nu^\alpha$; flat or slightly inverted, i.e. $-0.5\la \alpha\la 0.5$). During the hard-to-soft state transition, the continuous jet switches off (being quenched by a factor of $\ga 100$ in the radio band; e.g., \citealt*{Fender99, Coriat11, Russell11, Russell19}). Meanwhile, discrete bright knots/plasmoids are observed to move outward relativistically (e.g., \citealt*{Mirabel94, Hjellming95, Fender99, Yang10, Russell19}). These plasmoids are interpreted as the ``episodic jet'' or ``transient jet'', and their radio emission is optically-thin ($\alpha< -0.5$; steep spectrum)\footnote{For the spectral property of the radio emission, we use optically-thin/steep/$\alpha<-0.5$ interchangably, and optically-thick/flat/$-0.5\la\alpha\la0.5$ interchangably as well.} and highly-polarized (see e.g., \citealt*{Fender04, Fender09, Yuan09a, Zhang15} for summaries). In the soft state, the continuous jet is always quenched, while residuals of the episodic ejecta (created during the hard-to-soft state transition) may still exist (e.g., \citealt*{Brocksopp13}), possibly due to the interactions between the ejecta and the surrounding environment (e.g., \citealt*{Corbel02}).

A fundamental tool in investigating the disk--jet connection is to probe the correlation between radio (monochromatic, $\lr=\nu L_\nu$ at e.g. 5 GHz) and X-ray (integrated, i.e. $\lx = \int L_\nu d\nu$ in e.g. the 1--10 keV band) luminosities based on their quasi-simultaneous measurements in the hard state \citep{Corbel00, Corbel03, Corbel13, Gallo03, Gallo12, Gallo14, Furst15, Plotkin17, IZ18, Rodriguez19}. It it found that the radio and X-ray luminosities follow a tight non-linear correlation (hereafter RX correlation), $\lr\propto \lx^p$ with the slope index $p\approx 0.6\pm 0.1$ \citep{Corbel03, Corbel13}.\footnote{A slight offset in the normalization among the full and failed (defined as the case of no transition into the soft state) outbursts are observed in GX 339-4 \citep{Furst15}.} This represents the standard RX correlation (cf. the black dashed curve in Figure \ref{fig:rxcorr}). With the BH mass $\mbh$ introduced as a new factor, this correlation was later extended to include low-luminosity active galactic nuclei (AGNs) also, and it is renamed the fundamental plane of BH activity in logarithmic space (e.g., \citealt*{Merloni03, Falcke04, Wang06, Panessa07, Li08, Gultekin09, Gultekin19, Qian18, Li18}). With an emphasis on the slope $p$ instead of the dependence on $\mbh$, below for simplicity it will still be referred as the RX correlation.

Since the discovery, the empirical standard RX correlation is broadly established among a majority of sources. However, different correlation slope $p$ is reported in some specified systems, e.g. the radio-loud AGNs \citep{Wang06, Panessa07, Li08}, the narrow-line Seyfert 1 galaxies \citep{Yao18}, and the faint/quiescent AGNs (\citealt*{Yuan09b, Xie17}, but see \citealt*{Mezcua18}).\footnote{Quiescent BHBs seem to follow the standard RX correlation \citep{Gallo14, Plotkin17}, maybe because they are still not dim enough in these observations (e.g., discussions in \citealt*{Xie17}, and a hint in \citealt*{Dincer18}).} Even within its typical dynamical range, sources with clear deviations to the standard RX correlation are observed in both BHBs (so called ``outliers" for BHBs, e.g., \citealt*{Corbel04, Coriat11, Jonker12, Brocksopp13}) and AGNs (e.g., \citealt*{Bell11, King11, King13, Xie16b}). As demonstrated in the prototype BHB H1743--322 \citep{Coriat11} and low-luminosity AGN NGC 7213 \citep{Bell11, Xie16b}, these outliers likely follow a hybrid correlation, i.e. a steep $p\approx1.3$-1.4 branch at the bright $\lx$ regime (see the blue dot-dashed curve in Figure \ref{fig:rxcorr}) and $p\sim 0$ branch at moderate $\lx$ regime. Hint on the recover back to the standard correlation is also observed in H1743-322 when it is sufficiently weak in $\lx$ \citep{Coriat11}. We note that the existence of a new RX correlation track is also confirmed from statistics, i.e. through the data-cluster analysis method \citet*{Gallo12} find from a sample of 18 BHBs that, besides of the standard one, there exists another $p\approx 0.98$ correlation at the $\lx > 10^{36}\ergs$ regime.

Physically, the standard RX correlation provides strong evidence for a tight connection between the hot X-ray emitting source (usually a hot accretion flow), and the radio source (usually a continuous jet), and it is understood under the coupled accretion--jet model, where a scale-invariant jet model is considered (e.g., \citealt*{Heinz03, Merloni03, Heinz04, YC05, Xie16}). The physics behind the hybrid RX correlation, as well as its connection to the standard one, on the other hand, remain unclear. One promising solution is to attribute the change in the correlation slope $p$ to the change in the mode of hot accretion flows (rather than that in jet physics. See \citealt*{Xie16} and Sec.\ \ref{sec:model}). Alternative models for the $p\approx 1.3$-1.4 branch can be found in \citet*{MM14, Cao14, Qiao15}, and alternative models for the $p\sim 0$ branch can be found in \citet*{IZ18, Espinasse18}. For completeness, we note that the episodic jet, which is typically associated with the state transition, may have a different origin. Likely it relates to the formation and catastrophic disruption of the magnetic flux rope above the surface of accretion flow, see \citet*{Yuan09a} for details.

\begin{table*}
\begin{center}
\caption{Quasi-simultaneous JVLA and \swift/XRT observations of GRS 1739--278}\label{tab:obsdata}
\vspace{-0.2cm}
\begin{tabular}{ccccccccc}
	\hline
	\hline
	 \multicolumn{4}{c}{\underline{\hspace{3.cm}JVLA\hspace{3.cm}}}& \multicolumn{5}{c}{\underline{\hspace{4.2cm}\swift/XRT\hspace{4.2cm}}} \\
	Date (MJD) & $F_{5}$ (mJy)& $F_{8}$ (mJy) & Spectral Index $\alpha$ & ObsID & Date (MJD) & Exposure (s) & $F_{\rm X}$ (erg s$^{-1}$ cm$^{-2}$) & State \\
	\hline
57183.23&$0.19^{+0.04}_{-0.04}$&$0.11^{+0.03}_{-0.03}$&$-1.63^{+1.12}_{-1.12}$ &00033812002 &57183.11&869.61&$1.07_{-0.02}^{+0.02} \times 10^{-9}$ & Hard\\
57189.28&$0.21^{+0.03}_{-0.03}$&$0.11^{+0.02}_{-0.02}$ &$-1.79^{+0.89}_{-0.89}$&00033812008 &57189.08&984.61&$1.76_{-0.14}^{+0.14} \times 10^{-9}$ & Intermediate\\
57194.21&$1.45^{+0.05}_{-0.05}$&$1.32^{+0.05}_{-0.05}$ &$-0.28^{+0.17}_{-0.17}$&00033812013 &57194.27&619.58&$1.26_{-0.16}^{+0.16} \times 10^{-9}$ & Soft\\
57199.20&$0.26^{+0.03}_{-0.03}$&$0.12^{+0.03}_{-0.03}$&$-2.13^{+1.00}_{-1.00}$ &\nodata&\nodata&\nodata&\nodata & Soft?\\
57210.23&$0.15^{+0.03}_{-0.03}$&$0.17^{+0.02}_{-0.02}$&$0.30^{+0.58}_{-0.58}$ &00033812017&57210.03&1892.36&$1.24_{-0.06}^{+0.06} \times 10^{-10}$ & Hard\\
57212.25&$0.14^{+0.03}_{-0.03}$&$0.11^{+0.04}_{-0.04}$&$-0.72^{+1.43}_{-1.43}$ &00033812018&57212.69&1293.76&$8.06_{-0.63}^{+0.63} \times 10^{-11}$ & Hard\\
57214.22&$0.45^{+0.03}_{-0.03}$&$0.37^{+0.03}_{-0.03}$&$-0.57^{+0.35}_{-0.35}$ &00033812019&57214.22&1534.21&$5.75_{-0.70}^{+0.70} \times 10^{-11}$ & Hard\\
57217.16&$0.27^{+0.03}_{-0.03}$&$0.23^{+0.03}_{-0.03}$&$-0.42^{+0.44}_{-0.44}$ &00033812020&57216.16&1969.95&$9.41_{-1.31}^{+1.31} \times 10^{-12}$ & Hard\\
57218.13&$0.16^{+0.03}_{-0.03}$&$0.11^{+0.03}_{-0.03}$&$-1.10^{+0.95}_{-0.95}$ &00033812021&57218.42&2153.76&$4.37_{-0.93}^{+0.93} \times 10^{-12}$ & Hard\\
57222.14&$0.20^{+0.02}_{-0.02}$&$0.13^{+0.02}_{-0.02}$&$-1.25^{+0.76}_{-0.76}$ &00033812023&57222.02&2127.87&$5.34_{-0.84}^{+0.84} \times 10^{-12}$ & Hard\\
57226.17&$0.52^{+0.02}_{-0.02}$&$0.51^{+0.02}_{-0.02}$&$-0.05^{+0.18}_{-0.18}$ &00033812025&57226.20&935.22&$1.95_{-0.25}^{+0.25} \times 10^{-11}$ & Hard \\
57229.22&$0.58^{+0.03}_{-0.03}$&$0.54^{+0.02}_{-0.02}$&$-0.20^{+0.18}_{-0.18}$ &\nodata&\nodata&\nodata&\nodata & Hard\\
57234.16&$0.57^{+0.03}_{-0.03}$&$0.48^{+0.02}_{-0.02}$&$-0.50^{+0.25}_{-0.25}$ &00033812026&57234.06&3745.90&$1.44_{-0.06}^{+0.06} \times 10^{-10}$ & Hard\\
57236.08&$0.57^{+0.03}_{-0.03}$&$0.50^{+0.03}_{-0.03}$&$-0.37^{+0.23}_{-0.23}$ &00033812027&57236.12&3457.74&$1.84_{-0.12}^{+0.12} \times 10^{-10}$ & Hard\\
57238.08&$0.47^{+0.03}_{-0.03}$&$0.31^{+0.02}_{-0.02}$&$-1.20^{+0.48}_{-0.48}$ &00033812028&57238.03&3708.70&$2.15_{-0.08}^{+0.08} \times 10^{-10}$ & Hard\\
57240.13&$0.62^{+0.03}_{-0.03}$&$0.54^{+0.02}_{-0.02}$&$-0.42^{+0.21}_{-0.21}$ &00033812029&57240.76&3978.57&$2.81_{-0.14}^{+0.14} \times 10^{-10}$ & Hard\\
57242.11&$0.48^{+0.02}_{-0.02}$&$0.37^{+0.02}_{-0.02}$&$-0.73^{+0.31}_{-0.31}$ &00033812030 &57242.49&4111.97&$5.31_{-0.17}^{+0.17} \times 10^{-10}$ & Hard\\
57244.07&$0.26^{+0.02}_{-0.02}$&$0.23^{+0.02}_{-0.02}$&$-0.34^{+0.33}_{-0.33}$ &\nodata&\nodata&\nodata&\nodata & Intermediate?\\
57247.06&$0.04^{+0.02}_{-0.02}$&$0.02^{+0.01}_{-0.01}$ &$-1.51^{+2.07}_{-2.07}$&00033812032 &57246.82&1844.34&$1.57_{-0.14}^{+0.14} \times 10^{-9}$ & Intermediate\\
57273.98&$0.17^{+0.02}_{-0.02}$&$0.15^{+0.02}_{-0.02}$&$-0.28^{+0.44}_{-0.44}$ &00033812043&57273.15&1448.96&$1.65_{-0.07}^{+0.07} \times 10^{-10}$ & Hard\\
57276.02&$0.36^{+0.02}_{-0.02}$&$0.33^{+0.02}_{-0.02}$&$-0.21^{+0.24}_{-0.24}$ &00033812044&57276.21&1948.78&$6.11_{-0.40}^{+0.40} \times 10^{-11}$ & Hard\\
57280.06&$0.21^{+0.02}_{-0.02}$&$0.20^{+0.02}_{-0.02}$&$-0.13^{+0.36}_{-0.36}$ &00033812045&57279.27&2139.22&$2.83_{-0.24}^{+0.24} \times 10^{-11}$ & Hard\\
57280.97&$0.15^{+0.02}_{-0.02}$&$0.14^{+0.02}_{-0.02}$&$-0.14^{+0.50}_{-0.50}$ &00081764002&57280.93&914.03&$1.66_{-0.31}^{+0.31} \times 10^{-11}$ & Hard\\
57283.95&$0.15^{+0.02}_{-0.02}$&$0.14^{+0.02}_{-0.02}$&$-0.22^{+0.49}_{-0.49}$ &00033812046&57284.40&2162.19&$8.01_{-1.30}^{+1.30} \times 10^{-12}$ & Hard\\
57287.02&\nodata&$0.19^{+0.03}_{-0.03}$&\nodata&\nodata&\nodata&\nodata&\nodata & Hard\\
57310.97&$0.76^{+0.09}_{-0.09}$&$0.79^{+0.05}_{-0.05}$&$0.11^{+0.29}_{-0.29}$ &00033812052&57310.67&1813.58&$1.16_{-0.16}^{+0.16} \times 10^{-10}$ & Hard\\
57327.95&$0.73^{+0.10}_{-0.10}$&$0.74^{+0.05}_{-0.05}$&$0.06^{+0.29}_{-0.29}$ &\nodata&\nodata&\nodata&\nodata& Hard\\
57329.95&$0.87^{+0.10}_{-0.10}$&$0.79^{+0.06}_{-0.06}$&$-0.27^{+0.32}_{-0.32}$ &\nodata&\nodata&\nodata&\nodata& Hard\\
\hline
57656.00&$0.42^{+0.07}_{-0.07}$&$0.43^{+0.05}_{-0.05}$&$0.06^{+0.37}_{-0.37}$ &00033812056&57655.89&965.91&$1.90_{-0.12}^{+0.12} \times 10^{-10}$ & Hard\\
57660.95&$0.56^{+0.09}_{-0.09}$&$0.28^{+0.10}_{-0.10}$&$-1.57^{+0.95}_{-0.95}$ &00033812058&57661.07&748.76&$1.00_{-0.20}^{+0.20} \times 10^{-10}$ & Hard\\
57680.93&$0.36^{+0.02}_{-0.02}$&$0.41^{+0.02}_{-0.02}$&$0.28^{+0.16}_{-0.16}$ &00081979002&57681.08&1903.88&$2.64_{-0.32}^{+0.32} \times 10^{-11}$& Hard\\
57692.86&$0.60^{+0.02}_{-0.02}$&$0.55^{+0.02}_{-0.02}$&$-0.20^{+0.13}_{-0.13}$ &\nodata&\nodata&\nodata&\nodata & Hard\\
\hline
\end{tabular}
\end{center}
\small
{\bf Notes.} $F_5$ represents radio flux at 5.26 GHz (2015 Jun. 10 -- 2015 Nov. 3, i.e. before MJD 57330) or 4.70 GHz (2016 Sep. 25 -- 2016 Oct. 31, i.e. since MJD 57656). $F_8$ and $F_{\rm X}$ are the radio and X-ray fluxes, respectively, at 7.45 GHz and between 1 keV and 10 keV. 
%\tablerefs{ (1) xxx ...; (2) xxx}
\end{table*}

\begin{figure*}
\centering
\includegraphics[width=\linewidth]{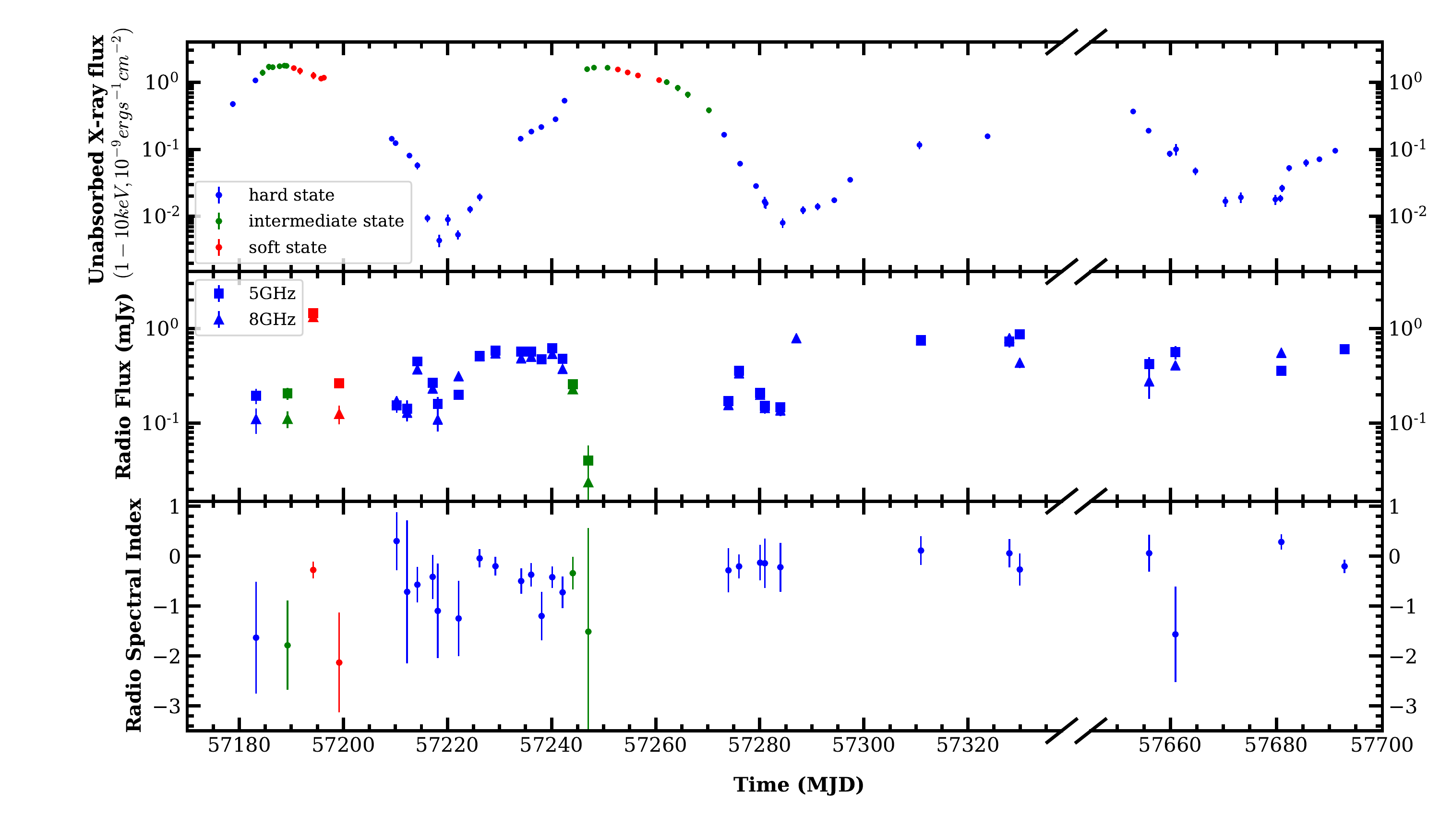}
\vspace{-0.2cm}
\caption{Light curves in X-rays (1-10 keV, top panel) and radio (middle panel). The bottom panel shows the spectral index $\alpha$ in radio. In all the panels, the accretion states are shown in color, i.e. the soft state in red, the hard state in blue, and the intermediate state in green. In the middle panel, the filed squares are at either 5.26 GHz (i.e. before MJD 57330) or 4.70 GHz (i.e. after MJD 57656), while the filed triangles are at 7.45 GHz. See Sec.\ \ref{sec:radioobs} for details.}
\label{fig:lc}
\end{figure*}

In this work we focus on the the 2015-2016 mini-outbursts of the X-ray transient GRS 1739--278. Located at a distance of 6-8.5 kpc \citep{Greiner96, Yan17b}, the binary system has an inclination of $i \approx 32.5\degr$ \citep{Miller15}. It was discovered in 1996 by SIGMA gamma-ray telescope on board the \textit{Granat} satellite \citep{Vargas97}, and is classified as a BH candidate based on the similarities in spectral and timing properties to other BHBs, as well as the detection of a strong quasi-periodic oscillation (QPO) in the intermediate state \citep{Borozdin00}. Eighteen years later, in 2014 GRS 1739--278 underwent a new main outburst \citep{Krimm14}, whose duration is remarkably long, i.e. more than one year. The peak X-ray luminosity of the 2014 outburst reaches $\sim5\times10^{38} \ergs$ \citep{Yan17b, Wang18}. Thanks to the intense X-ray monitoring of \swift\, afterwards, a series of mini-outbursts have been discovered, where state transitions are also detected in the first two mini-outbursts (see \citealt*{Yan17a, Yan17b} and also \autoref{fig:lc}).

This work is organized as follows. In Section \ref{sec:obs} we report the data analysis of these mini-outbursts, while in Section \ref{sec:results} we compile these data and investigate the radio properties. We find this source follows a flat RX correlation, for more than two orders of magnitude in the variation of X-ray luminosity. A discussion with a brief summary is given in Section \ref{sec:summary}. Throughout this work, the distance of GRS 1739--278 is fixed to $d=7.5$ kpc \citep{Yan17b}.

\section{Observation and Data Reduction}\label{sec:obs}

After a main outburst in 2014, GRS 1739--278 underwent a series of mini-outbursts. These mini-outbursts are fainter by a factor of $\ga 10$-30 to the main outburst in 2014, and have been monitored with a moderate cadence by \swift\, \citep{Yan17a, Yan17b, Parikh18}. As listed in detail in \autoref{tab:obsdata}, it is also covered by the Karl G. Jansky Very Large Array (JVLA) at 32 epochs (see Sec. \ref{sec:radio} later). Since we are mostly interested in the RX correlation, we only consider quasi-simultaneous observations, where the X-ray-to-radio time interval is required to be less than one day. With this quasi-simultaneity requirement, we only have 25 pairs of observations. 

\autoref{tab:obsdata} shows the details of all the observations. In the radio part we include the observation date (the modified Julian date, MJD) and fluxes at two wavebands. The spectral index in radio is also provided. In the X-ray part we include the observation ID and the date of \swift/XRT observations, the exposure time, the X-ray flux and the spectral state of the system. The exact time interval between radio and X-ray observations can then be derived easily.
%In \autoref{tab:obsdata}, the reported errors of the X-ray flux correspond to the 90\% confidence level. The errors of the radio flux include the root-mean-square in the image ($\sigma_{\rm rms}$) as well as a 3\% uncertainty in the absolute flux density scale, following \citet{Nyland17}.}

\subsection{\swift/XRT Observation and Spectral Analysis} \label{sec:xray}

The data reduction and analysis in X-rays are done through standard procedures, i.e. the \swift/XRT event data were first processed with {\scriptsize XRTPIPELINE} (v 0.13.2) to generate the cleaned event data, and then grade 0 events are extracted by {\scriptsize XSELECT}. For those with high count rate ($>0.6$ counts s$^{-1}$ for the photon counting mode, and $>150$ counts s$^{-1}$ for the windowed timing mode), the events in the central region that suffer pile-up effect are excluded. The details can be found in \citet{Yan17b}. Below we provide a brief description on the spectral modeling.

The X-ray spectrum is modeled by an absorbed power-law component from the hot accretion flow, and a thermal component from cold disk, i.e. {\it tbabs*(powerlaw+diskbb)} in {\scriptsize XSPEC} notation. The hydrogen column density $N_H$ of the absorption in soft X-rays is constrained to be $\approx 2.5\times10^{22}$ cm$^{-2}$, which is in good agreement with individual measurements by \xmm\,  and \nustar\,  \citep{Miller15, Furst16}. The spectral state is of crucial importance for our investigation. We follow \citet*{RM06, Belloni10} to define the hard state if the emission is dominated by a power-law component whose X-ray photon index is also less than 2.1, the soft state as the flux contribution from disk thermal component being larger than 80 per cent, and the in-between spectra are defined as in the intermediate state (see also \citealt*{Yan17a}). Note that thanks to the intense monitoring in X-rays, 25 epochs out of total 32 radio observations have quasi-simultaneous X-ray observations, thus their accretion states can be clearly determined. For the rest (7/32) radio epochs that lack the X-ray information, their accretion state is constrained/estimated based on the evolutionary trend as well as X-ray observations at two adjacent epochs (epochs before and after the radio observation). Take MJD 57229 as an example. X-ray monitoring indicates that GRS 1739--278 is in the hard state during the period of MJD 57220--57243 (more specificlly, on MJD 57226 and MJD 57234, see \autoref{fig:lc} and \citealt*{Yan17a}). We thus argue that GRS 1739--278 is also in the hard state on MJD 57229, as listed in \autoref{tab:obsdata}.

We calculate the 1-10 keV X-ray flux from the spectral fitting by the model {\it cflux}, which are shown in the top panel of \autoref{fig:lc} (and \autoref{tab:obsdata}), where uncertainties in the flux are evaluated at 90\% confidence. At least five mini-outbursts are captured by \swift/XRT, among which the first two have monitoring in both the rise and the decay phases. The dynamical range in X-rays is almost three orders of magnitude, between $\sim4\times10^{-12}\ergscm$ and $\sim2\times10^{-9}\ergscm$. Even for those hard state with radio data only, the dynamical range in X-rays is still more than two orders of magnitude.

\subsection{JVLA Radio Observation and Data Reduction} \label{sec:radioobs}

\begin{figure*}
\centering
%\vspace{-8cm}
\includegraphics[width=0.3\linewidth]{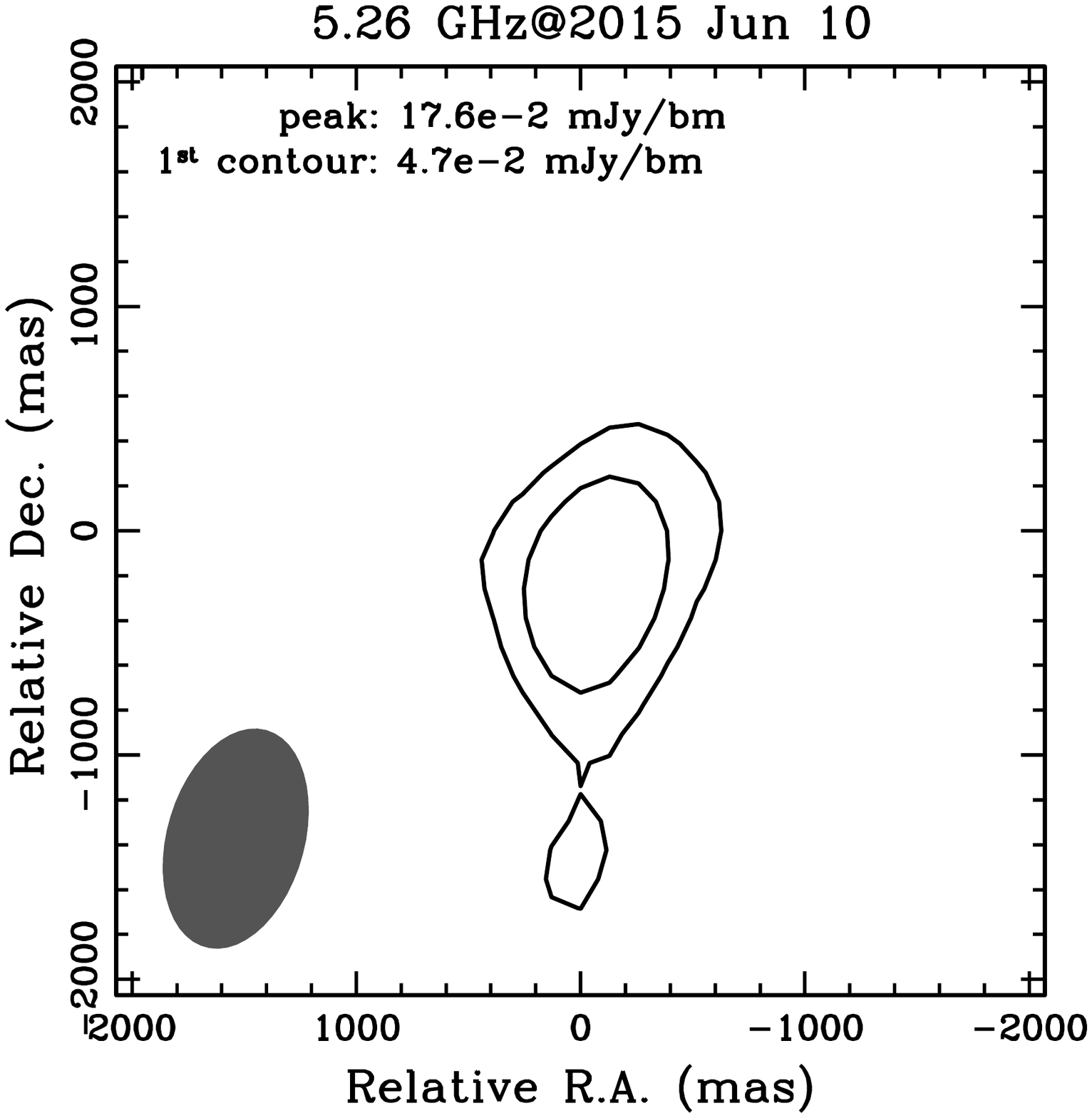}
\hspace{-0.5cm}
\includegraphics[width=0.3\linewidth]{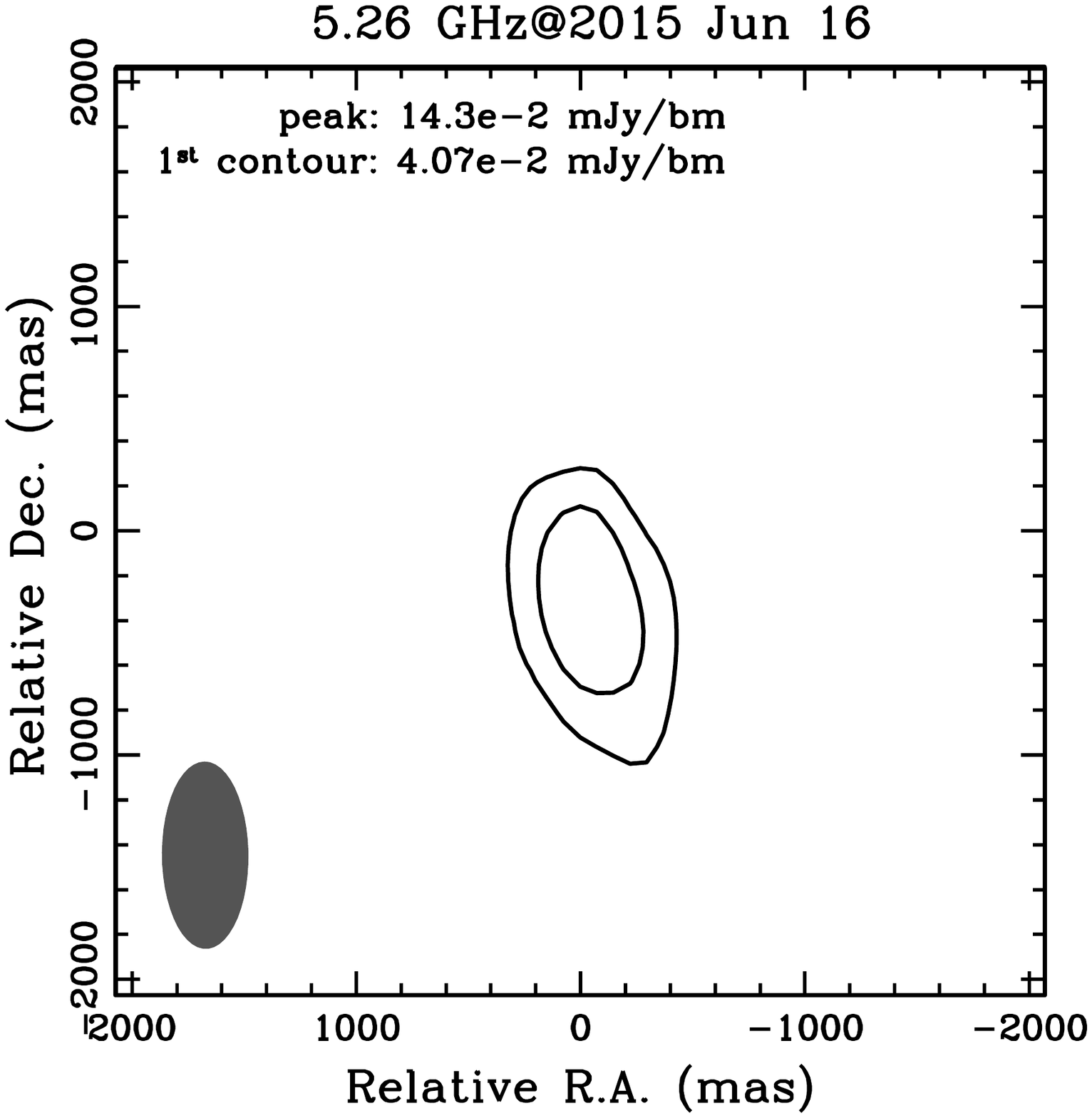}
\hspace{-0.5cm}
\includegraphics[width=0.3\linewidth]{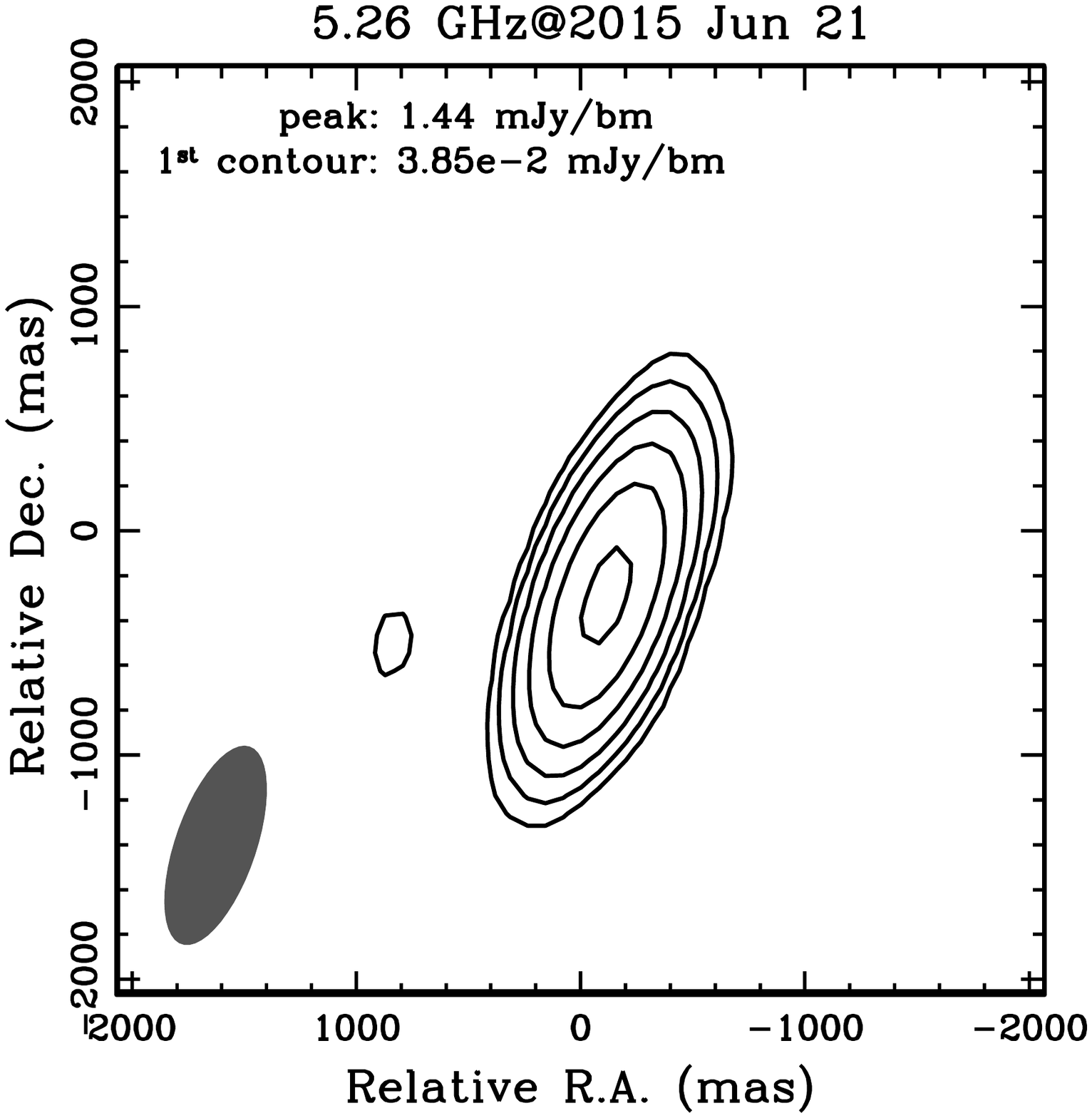}\\
\vspace{-2.5cm}
\includegraphics[width=0.3\linewidth]{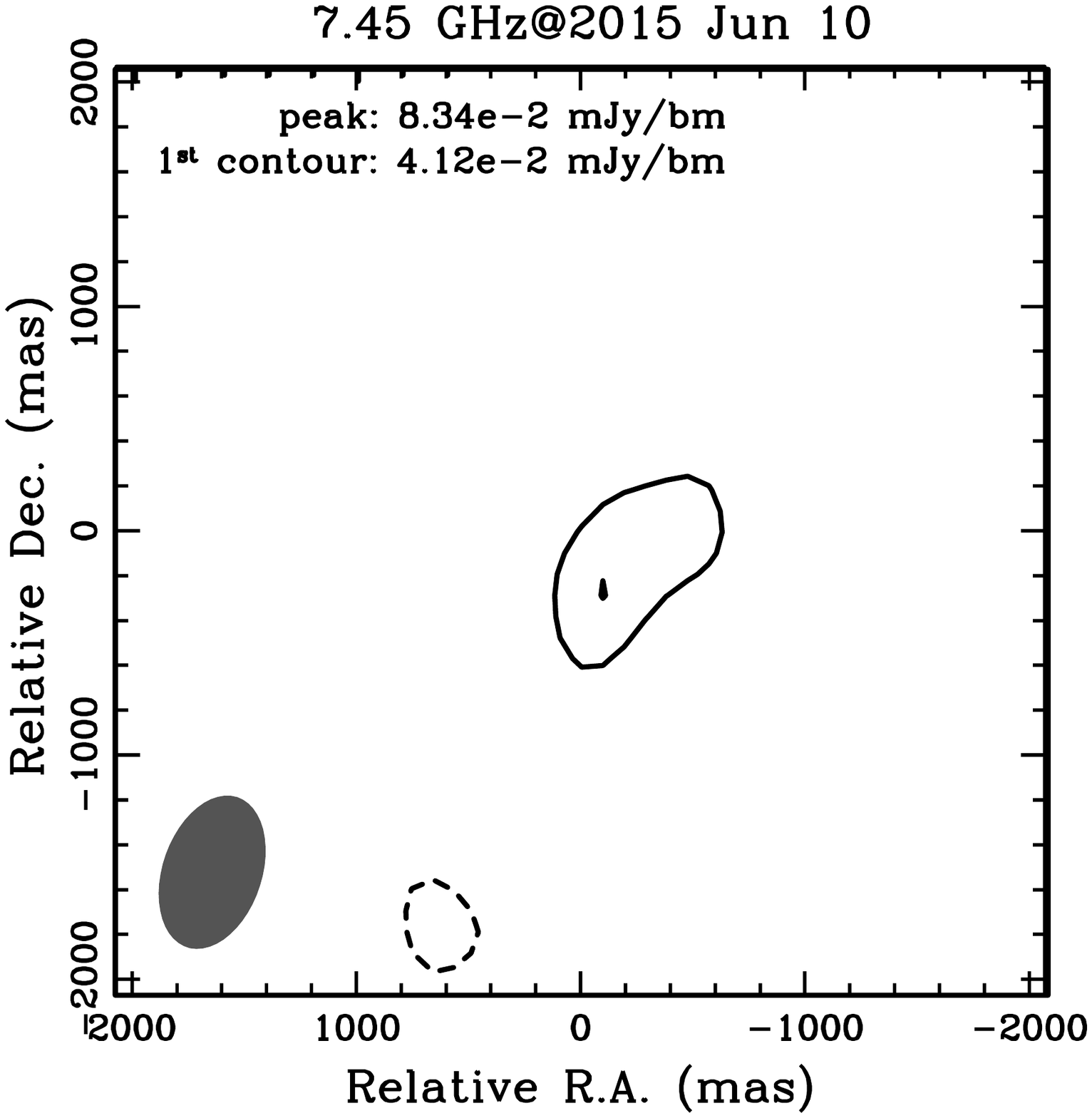}
\hspace{-0.5cm}
\includegraphics[width=0.3\linewidth]{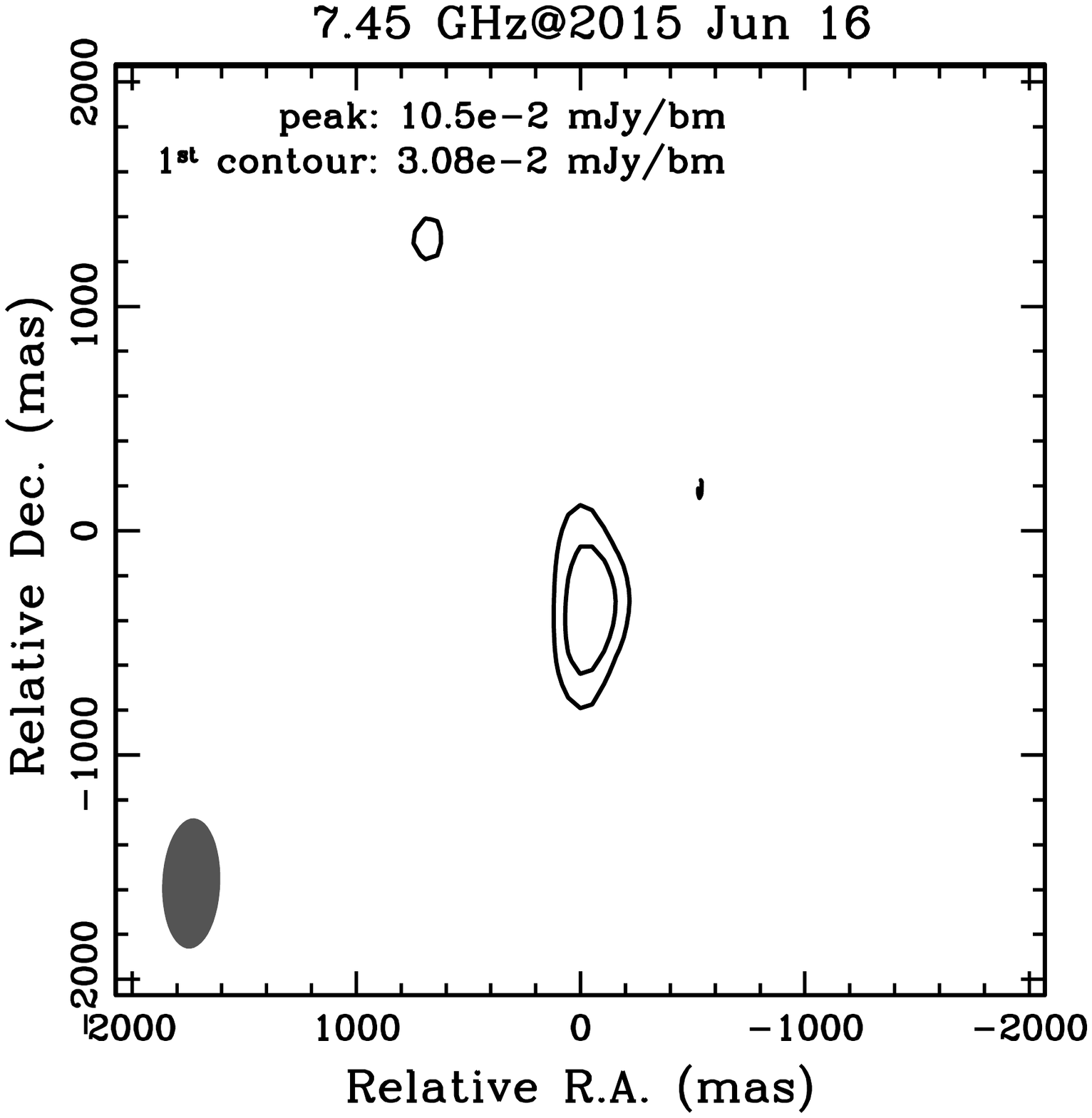}
\hspace{-0.5cm}
\includegraphics[width=0.3\linewidth]{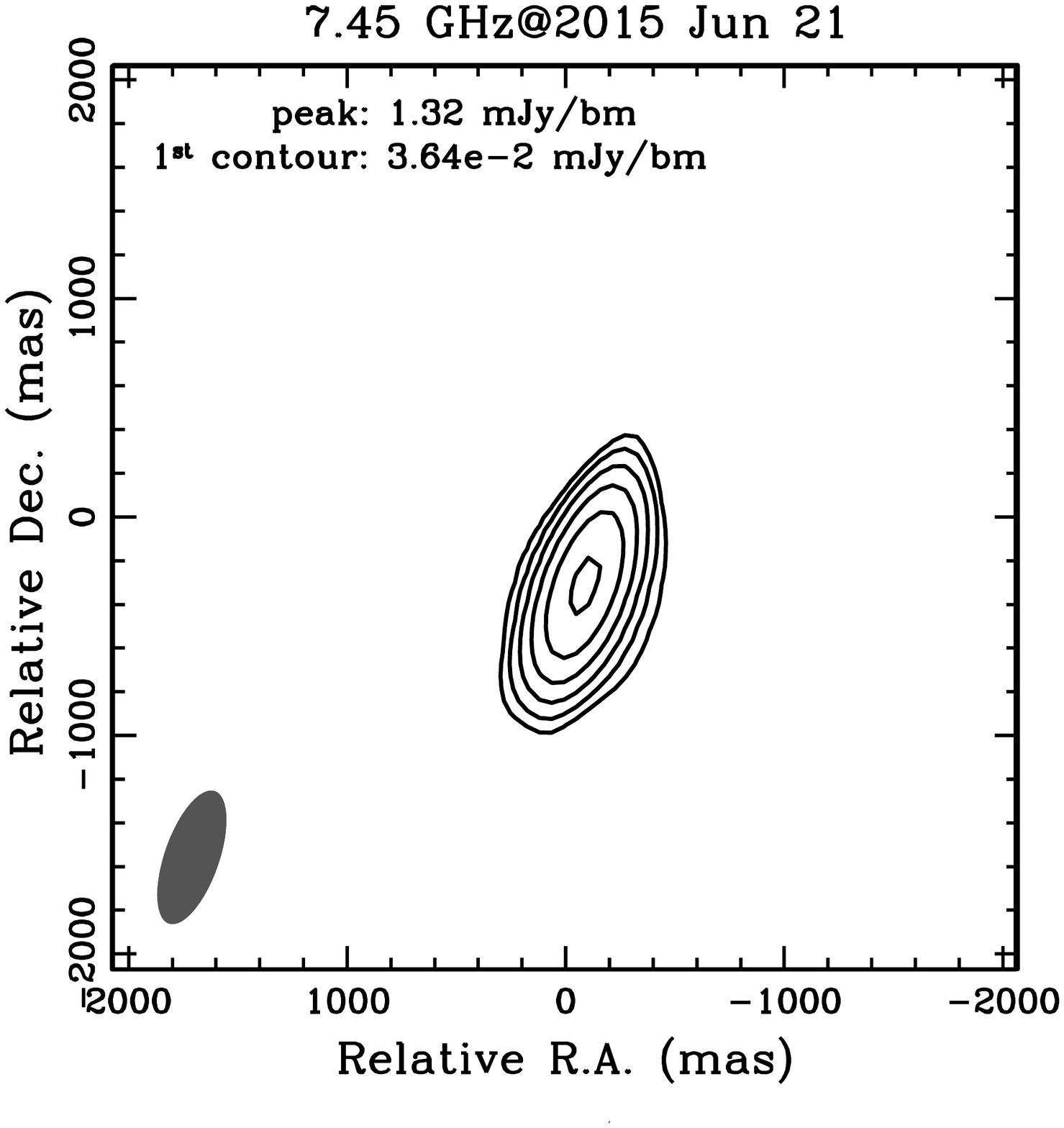}
\vspace{-1.5cm}
\caption{JVLA radio image of GRS 1739--278 at 5.26 GHz (top panels) and 7.45 GHz  (bottom panels). The plots center at R. A. = 17$^{\rm h}$42$^{\rm m}$40.030$^{\rm s}$ and Dec. = -27$\degr$44$'$52.699$''$. From left to right, they are respectively on 2015 Jun. 10 (MJD 57183, hard state), 2015 Jun. 16 (MJD 57189, intermediate state) and 2015 Jun. 21 (MJD 57194, soft state). The shadow in each panel shows the beam size (FWHM). The peak flux per beam (``bm'' in the figure) as well as the the first (solid curve) contour are labeled in each panel. The contours increases by a factor of 2, i.e. they follow (-1, 1, 2, 4, 8, ...).}
\label{fig:radioimage}
\end{figure*}

Radio observations of GRS 1739--278 were obtained by JVLA (Project code VLA/SB4161 and SH0281, PI: S. Corbel) between 2015 June 10 and 2016 October 31, with a total of 32 epochs. The time interval between two neighboring radio epochs varies, with a typical value of $\approx$2.5-5 days during the first two mini-outbursts (cf. \autoref{tab:obsdata} and \autoref{fig:lc}). It is observed at C band, simultaneously centering at two broad frequencies (hereafter subbands). One subband centers at 7.45 GHz, and the other centers at either 5.26 GHz (2015 Jun. 10 -- 2015 Nov. 3, i.e. before MJD 57330) or 4.70 GHz (2016 Sep. 25 2016 -- 2016 Oct. 31, i.e. since MJD 57656). Both subbands have a bandwidth of 1 GHz, i.e. the total bandwidth is 2 GHz. The on-source exposure time varies between 18 and 80 minutes. The JVLA observations are in the A configuration, with a typical spatial resolution (full-width-half-maximum, FWHM) of about 900 mas $\times$ 350 mas at 5 GHz and of about 650 mas $\times$ 250 mas at 7.45 GHz (mas is the abbreviation of milliarcsecond; see \autoref{fig:radioimage}).

%more accurately, 922  mas$\times$357 mas at 5 GHz (ZZWu). 

The calibration was performed using the standard JVLA pipeline of the Common Astronomy Software Application v5.4.1 ({\scriptsize CASA}, \citealt*{McMullin07}). After calibration, the data of target sources is split and exported out (in fits format). Further processes including the imaging are done in {\scriptsize Difmap} software \citep{Shepherd94}. Depending on exposure time and wavelength, the $1\sigma_{\rm rms}$ sensitivity we achieved varies between $\approx 6\,\mu$Jy beam$^{-1}$ and $\approx 50\,\mu$Jy beam$^{-1}$. Finally, the flux density at each epoch is derived by fitting a point source in the image plane using the task {\scriptsize MODELFIT}, and the results are summarized in \autoref{tab:obsdata}. We note that for the 5.26 GHz observation on 2015 Sep. 22 (MJD 57287), the whole calibrated data is flagged out after the pipeline reduction, and is thus not considered in our data analysis. For the uncertainties of the radio flux reported in \autoref{tab:obsdata}, we follow \citet*{Nyland17} to include both $\sigma_{\rm rms}$ of the image and a systematic 3\% uncertainty in the absolute flux density scale, i.e. we have $\sigma_{\rm tot} = \sqrt{(2\,\sigma_{\rm rms})^2 + (0.03\,S_{\rm peak})^2}$, where $S_{\rm peak}$ is the peak intensity (flux per beam).

Once the fluxes at two frequencies are measured, we can then evaluate the spectral index $\alpha$ (see the bottom panel of \autoref{fig:lc}), where the uncertainty in the frequency because of the broad 1 GHz bandwidth is also taken into account.

\begin{figure}
\centering
\includegraphics[angle=-90, width=1.1\linewidth]{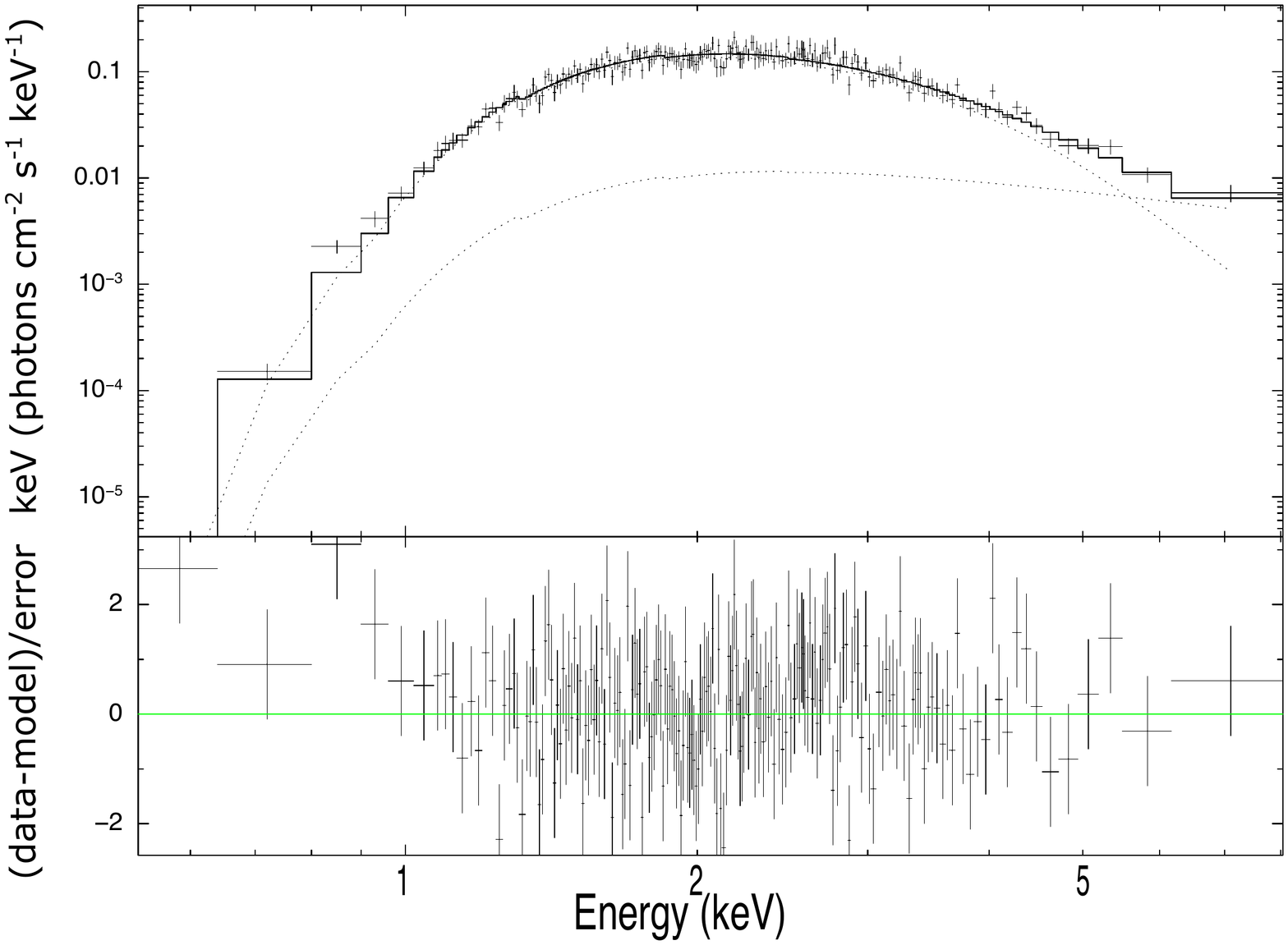}
\caption{The spectral modeling of the X-ray spectrum of GRS 1739--278 on MJD 57194. In the upper panel, the two dotted curves are the absorbed emission of {\it diskbb} (upper one at 5 keV) and {\it powerlaw} (lower one at 5 keV), and the solid curve represents the sum.}
\label{fig:soft}
\end{figure}

\section{Results} \label{sec:results}

The X-ray lightcurve of GRS 1739--278 of the 2015--2016 mini-outbursts is shown in the top panel of \autoref{fig:lc}. The X-ray properties are analyzed and discussed in detail in \citet*{Yan17b}. As shown in the top panel of \autoref{fig:lc}, state transitions, typically observed in the main outbursts, are also observed in the first two mini-outbursts. Note that \citet*{Yan17b} found that the much-fainter soft state in mini-outbursts likely follows the same tight $L_{\rm bol}\propto T^4$ (bolometric luminosity $L_{\rm bol}$ and representative temperature of the cold disk $T$) relationship as determined by that in the major 2014 outburst, suggesting that the cold disk is also not truncated in these soft states of mini-outbursts.

Below we focus on the radio observations and the disk--jet coupling.

\begin{figure*}
\centering
\includegraphics[width=0.6\linewidth]{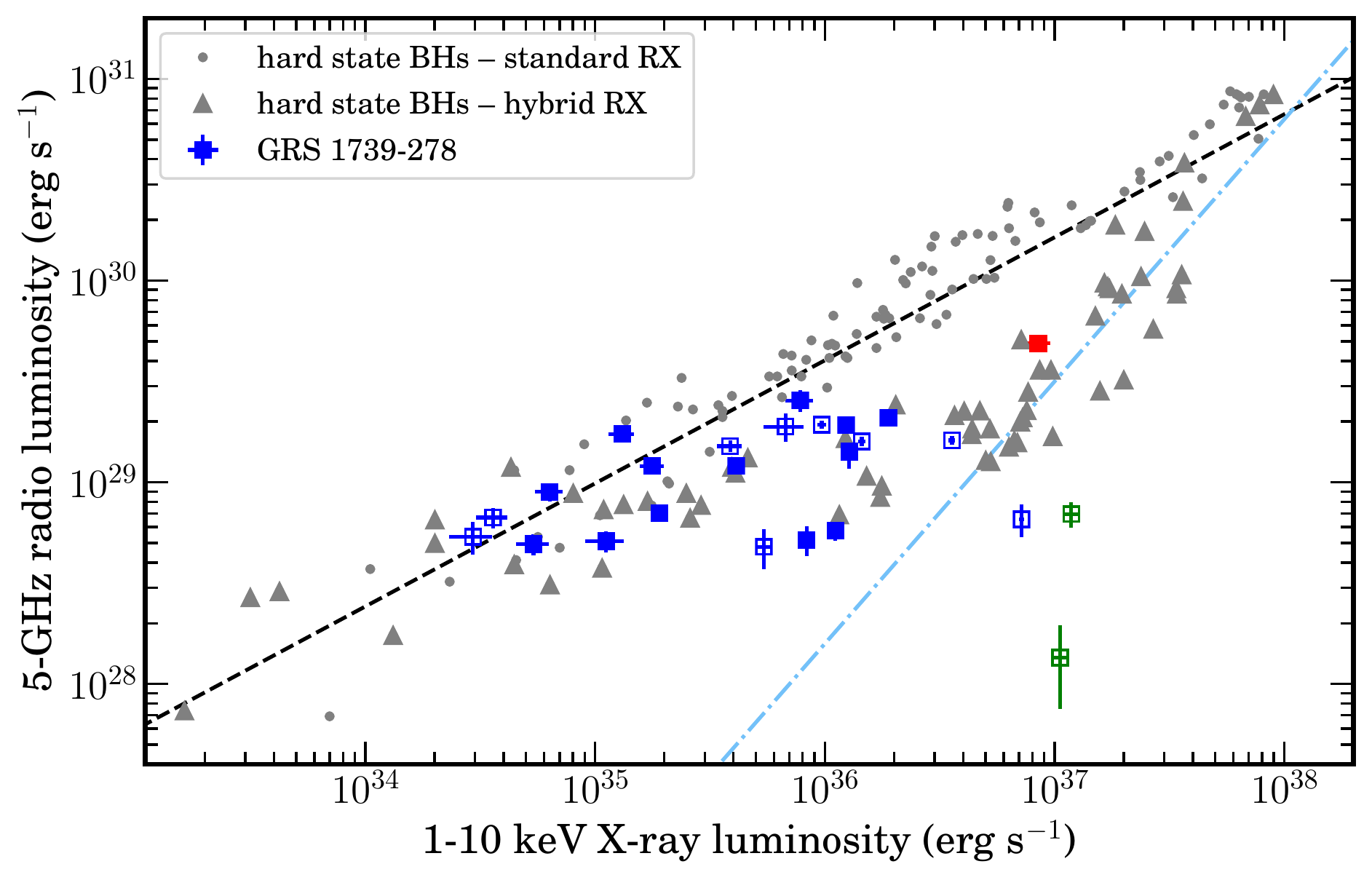}
\caption{The radio/X-ray correlation. The squares show the data of GRS 1739--278, with color follows \autoref{fig:lc}. Here the filled ones are optically-thick and the open ones are optically-thin. Data of other BHBs, compiled by \citet*{Bahramian18}, are shown by grey filled circles and triangles, respectively, for standard ($p\approx 0.61$, see \citealt*{Corbel13} and the black dashed curve for the fitting) and hybrid (see \citealt*{Coriat11}, and blue dot-dashed for the $p\approx 1.3$ branch) RX correlations. The colors of data points of GRS 1739--278 are the same to those of \autoref{fig:lc}.}
\label{fig:rxcorr}
\end{figure*}

\subsection{Image and Spectral Properties in Radio} \label{sec:radio}

The spatial morphology of GRS 1739--278 remains compact (i.e. unresolved) among almost all our A-configuration VLA observations. For illustrative purpose we show in \autoref{fig:radioimage} the image at 5.26 GHz (top panels) and 7.45 GHz (bottom panels) in three different states, i.e. in hard state on 2015 Jun. 10 (MJD 57183, left panels), in intermediate state on 2015 Jun. 16 (MJD 57189, middle panels) and in soft state on 2015 Jun. 21 (MJD 57194, right panels). In all the plots, the central position locates at R. A. = 17$^{\rm h}$42$^{\rm m}$40.030$^{\rm s}$ and Dec. = -27$\degr$44$'$52.699$''$ in the J2000 coordinate, which is determined by the VLA observation (in C configuration) of its 1996 outburst \citep{Durouchoux96}. We do not observe any offset in the position (in neither R.A. nor Dec.) in any spectral states, even in the intermediate and soft states where episodic ejections with superluminal motions are commonly observed (for superluminal ejections, see e.g., \citealt*{Mirabel94, Hjellming95, Fender99, Yang10, Russell19}). 

Besides, as shown in \autoref{fig:radioimage}, we find that the radio location of GRS 1739--278, determined by the JVLA observations in A configuration, whose spatial resolution is higher than that in C configuration, systematically shifts by 78.5 mas in R. A. and -313.9 mas in Dec., i.e. the actual position of GRS 1739--278 determined by JVLA on 2015 Jun. 21 (brightest in radio) is 
\begin{eqnarray}
{\rm R. A. (J2000)} & = & 17^{\rm h}42^{\rm m}40.022^{\rm s} \pm 0.002^{\rm s}, \nonumber\\
{\rm Dec. (J2000)} & = & -27\degr44'52.981''\pm0.005''.
\end{eqnarray}
Here we only include the statistical errors on the fit of beam centroiding, i.e. evaluated as beam/2$\times S_{\rm peak}/\sigma_{\rm rms}$. 

The middle and bottom panels of \autoref{fig:lc} show respectively the lightcurve and the spectral index in radio. Consistent with other BHBs in hard state, the radio spectrum of the hard state in GRS 1739--278 is typically-thick. There are several hard-state epochs whose radio spectrum seems steep (optically-thin), i.e. on 2015 Jun. 10 (MJD 57183, see also the left panels in \autoref{fig:radioimage} for radio images) and on 2016 Sep. 30 (MJD 57661). If real, they may possibly relate to the episodic ejections in the hard state \citep{Yuan09a} or the quiescent state (e.g., in our Galaxy center Sgr A*, \citealt*{Dodds-Eden11}). However, as shown in \autoref{fig:lc}, the value of $\alpha$ at these epochs is not firmly measured, we thus avoid further discussions.

\subsection{Radio Evolution in Intermediate and Soft States} \label{sec:im_soft}

We here focus on the radio evolution in intermediate and soft states. The radio evolution in hard state will be addressed subsequently in Sec.\ \ref{sec:rxcorr}.

Due to the sparse schedule of the radio monitoring, we unfortunately did not catch the jet evolution in the intermediate state during the rise phase of the first mini-outburst (around MJD 57180-57190). Instead, we have two epochs of intermediate state observations during the rise phase of the second mini-outburst (around MJD 57240-57250; see \autoref{fig:lc}), where we likely observe an un-finished jet quenching process within 5 days, i.e. with an increase in X-ray flux by a factor of $\sim 3$, the radio flux reduces by a factor of $\sim 12$ at 5.26 GHz and $\sim 18$ at 7.45 GHz, from MJD 57242 to MJD 57247 (i.e. during the hard-to-soft state transition). Meanwhile, the radio spectrum also steepens, consistent with transient ejections. We note the jet quenching during the state transition is commonly observed in main outbursts \citep[e.g.][]{Fender99, Coriat11}, and it has also been observed during the state transition of mini-outbursts in another BH transient MAXI J1535--571 \citep{Parikh19}.

Among all the 32 epochs of radio observations, there is only one epoch, i.e. on 2015 Jun. 21 (MJD 57194), that the system is in the soft state.\footnote{The epoch on 2015 Jun. 26 (MJD 57199) is a soft state candidate, where a steep radio spectrum ($\alpha =-2.13\pm 1.0$) is observed. But without X-ray observation the spectral state cannot be confirmed (cf. \autoref{tab:obsdata}).} The X-ray observations show that this source enters into the soft state about 3 days ago, on MJD 57191 (more accurately, between MJD 57190.48 and MJD 57191.67, see \autoref{fig:lc} and \citealt*{Yan17b}.). As shown in \autoref{fig:soft}, the X-ray spectrum on MJD 57194 is well-fitted under the adopted model, where the disk component contributes $\approx88\%$ of total flux in 1-10 keV. The best-fit values of the inner disk temperature and the photon index are $T_{\rm{in}} = 0.66^{+0.03}_{-0.01}$ keV and $\Gamma=2.13^{+0.93}_{-0.44}$, respectively. All these properties justify our soft state classification on MJD 57194. In this soft state (on MJD 57194) the system reaches maximal radio flux among all the 32 radio observations. Besides, the radio emission is spatially unresolved (see right panels of \autoref{fig:radioimage}) and the radio spectrum is marginally optically-thick, with $\alpha\approx -0.28\pm 0.17$ (cf. \autoref{fig:lc}).

\begin{figure}
\centering
\includegraphics[width=0.95\linewidth]{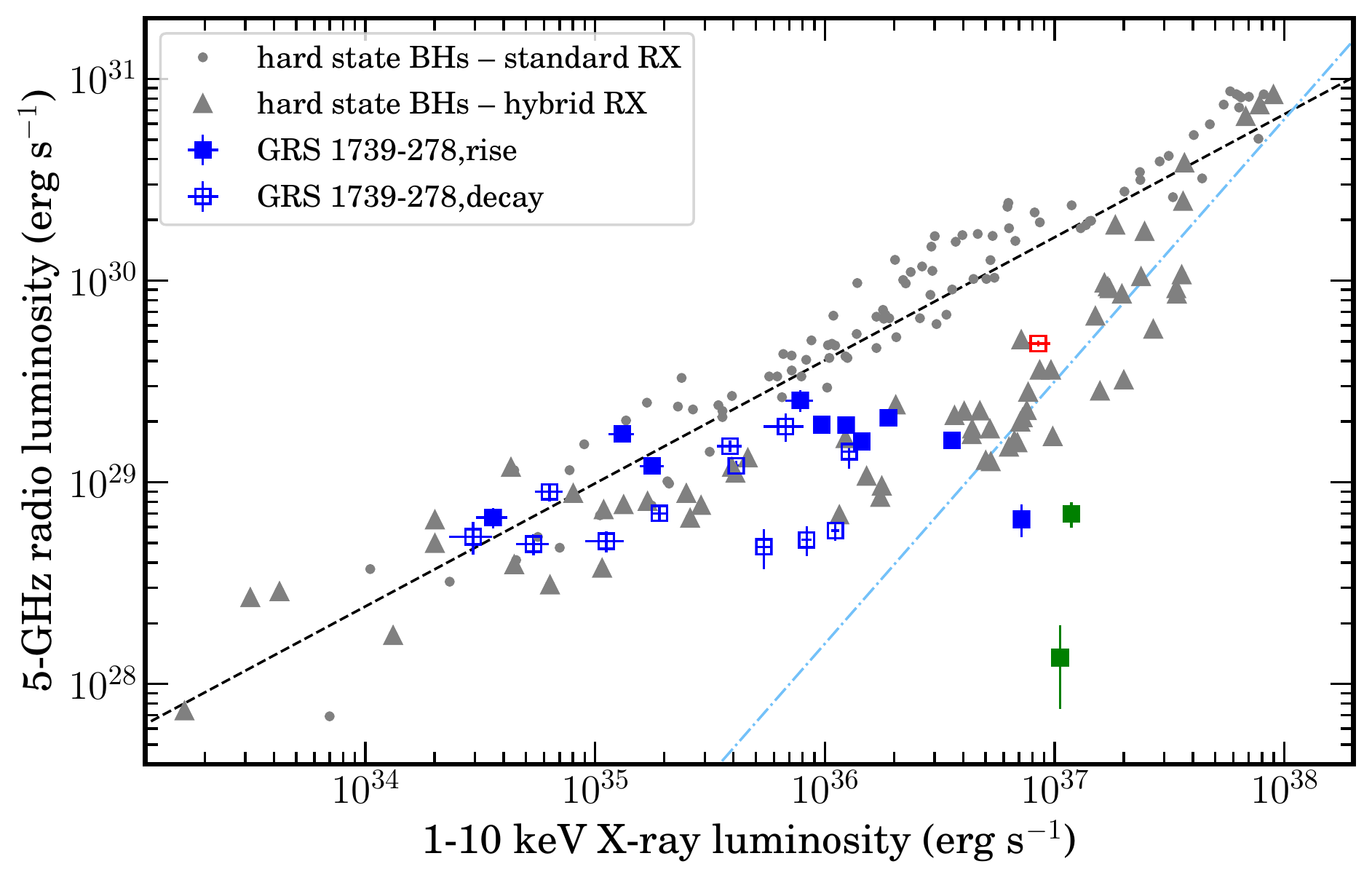}
\caption{The radio/X-ray correlation. This plot is the same as \autoref{fig:rxcorr}, except that the rise and decay phases of GRS 1739--278 are shown separately by filled (rise phase) and open (decay phase) squares.}
\label{fig:rxcorr2}
\end{figure}

\subsection{Radio/X-ray Correlation in Hard State} \label{sec:rxcorr} 

We now investigate the relationship between luminosities in 5 GHz radio and 1-10 keV X-rays for GRS 1739--278. As shown in \autoref{fig:rxcorr}, observations in hard state are focused. In this plot, the 2015-2016 mini-outbursts are shown by blue squares, where optically-thick (spectral index clusters around $\alpha \approx -0.2$) and optically-thin ($\alpha$ clusters around $\alpha \sim -1$) data points are shown respectively in filled and open symbols. We also show in this plot data points in soft (red squares) and intermediate (green squares) states of GRS 1739--278. For comparison, we also show the RX correlation of other BHBs in their hard states, where the data are taken from the latest compilation of \citet*{Bahramian18}.\footnote{https://github.com/bersavosh/XRB-LrLx\_pub} Sources that follow the standard $p\approx 0.61$ correlation (\citealt*{Corbel13}, the black dashed curve for the fit) are shown by black filled circles, and those that follow the hybrid correlation (\citealt*{Coriat11}, the blue dot-dashed curve for the fit of the $p\approx 1.3$ branch) are shown by the black filled triangles.

Several results can be derived immediately. First, we find that for GRS 1739--278, there is no clear difference between optical-thick jets and optical-thin ones. Secondly, the radio flux varies with a rather large scatter at a given X-ray luminosity, with a weak hint on the existence of two tracks, one is systematically fainter than the other by a factor of $\sim 4$ at given X-ray luminosity. However, the evidence of two tracks is weak. Below we omit this separation. 

Thirdly, apart of the large scatters, the RX relationship in GRS 1739--278 shows a clear deviation to the standard $p\approx 0.61$ one. We run a linear fit between $\log\lr$ and $\log\lx$ for all the data points in hard state (including both optical-thick ones and optical-thin ones) and the result is
\begin{eqnarray}
\log(\lr/\ergs) & = & (0.16\pm 0.09)\,\log(\lx/\ergs) \nonumber\\ 
& &+ 23.3\pm 3.1, \label{eq:rxcorr}
\end{eqnarray}
i.e. for a range of more than two orders of magnitude in X-ray luminosity this source exhibits a rather flat RX correlation. Interestingly, H1743--322 also follow a flat correlation in this X-ray luminosity regime (transition regime in the notation of \citealt*{Coriat11}), and the dynamical range in X-ray luminosity of this flat correlation branch is also similar to our results, cf. \citet*{Coriat11}.

For a detailed RX correlation investigation of GX 339-4, \citet{Furst15} find that the rise and decay phases of outbursts follow different tracks (see also \citealt*{IZ18}). To examine this possibility in GRS 1739--278, we in \autoref{fig:rxcorr2} separate the rise and decay phases by the solid and open squares, respectively, Note that because of the hysteresis phenomenon, the X-ray luminosity of the rise phase is systematically brighter than that of the decay phase. Except of this difference, there is no clear difference in the RX correlation among these two phases, especially when the large scatters in the data points are considered.

\section{Summary and Discussions} \label{sec:summary}

\subsection{Summary}
In this work, we analyzed the JVLA radio observations of the 2015-2016 mini-outbursts of GRS 1739--278. The JVLA monitoring campaign has simultaneous detections at 5 GHz and 8 GHz, and runs from 2015 Jun. 10 till 2016 Oct. 31. Among the 32 epochs, 25 epochs have quasi-simultaneous X-ray observations by \swift/XRT within one day (see \autoref{tab:obsdata}). The position of GRS 1739--278 constrained by our JVLA observation in A configuration is R. A. = 17$^{\rm h}$42$^{\rm m}$40.022$^{\rm s}$ and Dec. = -27$\degr$44$'$52.981$''$ in the J2000 coordinate.

The main results of this work can be summarized as follows,
\begin{itemize}
\item The radio image of GRS 1739--278 remains unresolved in all our A-configuration JVLA observations, whichever state it is and whatever spectral properties it has in radio band. No superluminal motion is observed in this source.
\item A majority of radio observations in the hard state show an optically-thick spectrum, consistent with previous findings (see \citealt*{Fender04, Fender09, Fender14} for reviews). Occasionally the radio spectrum in the hard state becomes optically thin, and the spectrum can be as steep as $\alpha \sim -1$, but admittedly the uncertainties in $\alpha$ are large. 
\item The jet quenching process is possibly caught during the intermediate state of the rise phase of the outburst, which represents an ongoing stage of hard-to-soft transition. On the other hand, we also spot an optically-thick ($\alpha\approx -0.28\pm 0.17$) radio emission in the soft state, which turns out to be the brightest among all the JVLA epochs.
\item For the RX correlation during the hard state, there is no clear difference between optical-thin jets and optical-thick ones. Moreover, for more than two orders of magnitude in the variation of X-ray luminosity, GRS 1739--278 follows a flat RX correlation with $p\approx 0.16\pm0.09$. Both the correlation slope and the X-ray luminosity regime agree well with the hybrid RX prototype H1743--322 \citep{Coriat11}, although neither the $p\approx 1.3$ correlation branch at the bright $\lx$ part nor the recover to the standard correlation branch at the faint $\lx$ part are observed in this source. Coordinated monitoring campaign in radio and X-rays of main and mini outbursts in future is in demand to examine the connection (and possibly the difference) among these two sources.
\end{itemize}

\subsection{Radio Emission in the Soft State}\label{sec:radiosoft}

Only on 2015 Jun. 21 the system is confirmed to be in the soft state, during which a very bright but marginally optically-thick ($\alpha\approx -0.28\pm0.17$) radio emission is detected. It remains unresolved in JVLA (A-configuration).

The radio emission in soft state, if observable, usually exhibits a steep optically-thin spectrum (see e.g., \citealt*{Fender04, Fender99, Yuan09a, Russell19}), which is produced by the residual episodic jet that launched during the hard-to-soft state transitions (see {\it Introduction}). It can be fairly bright if the ejecta interacts with dense medium (e.g., with stellar wind in Cyg X-3, see e.g., \citealt*{Koljonen10}). In GRS 1739--278 such optical-thin radio emission in the soft state is indeed observed on MJD 57199 (see Sec. \ref{sec:radio}) and during the 1996 main outburst \citep{Hjellming97}. The soft state of the 1996 outburst, crudely identified from the hardness ratio measurement in X-rays by \rxte/ASM, lasts over 170 days (between MJD 50166 and MJD 50338). VLA observations reveal that the radio emission is always optically-thin during this period \citep{Hjellming97}. 

We note that, in the episodic ejection model a flat spectrum can be observed at the early phase, when the ejecta is small and dense, thus synchrotron self-absorption peaks at a higher frequency. However, this implies that the radio emission peaks on some day later. Considering the low radio flux 5 days later on MJD 57199 (i.e. the candidate soft state), an efficient cooling of the relativistic electrons and/or decline in magnetic field strength in the ejecta will be necessary. For the soft state radio observation on MJD 57194 (2015 Jun. 21), we are thus disfavor the episodic ejection during the state transition, but instead in favor of the interpretation of an optically-thick emission from continuous jet.

Among all the BHBs discovered so far, to our knowledge the only other source with continuous jet in soft state is Cyg X-3 \citep{Zdziarski18}, where a radio-to-X-ray time lag of $\sim 50$d is also observed. However, Cyg X-3 is a well-known high-mass X-ray binary, and the accumulation of magnetic flux supplied by the high-mass companion in the soft state is argued to be crucial for launching of jet in soft state \citep{Cao20}. This mechanism cannot operate in low-mass binaries like GRS 1739--278 \citep{Cao20}.

Our detection of the optically-thick radio emission during the soft state may imply that, we possibly find in BHBs the counterpart of radio-loud active galactic nuclei (and quasars) that also have continuous jets. These systems are high in Eddington ratio $L_{\rm bol}/\ledd$ (Eddington luminosity for a black hole with mass $\mbh$ is, $\ledd=1.3\times10^{46}(\mbh/10^8 \msun)\ergs$), and are dominated by thermal emission from cold disk, as the case of soft state in BHBs. Intense coordinated monitoring in radio and X-rays during the soft state in future are necessary to verify it.

\subsection{Theoretical Interpretation of Radio/X-ray Correlation}\label{sec:model}

The physical reason for the hybrid RX correlation, as well as its connection to the standard one, remain poorly understood. Several scenarios have been proposed in literature. Below we examine our data with these existing models.

The first scenario of the hybrid RX emphases on the difference between the rise and the decay phases of the outburst. \citet*{IZ18} analyzed the evolution of H1743--322 and GX 339-4, where they reported that the different branches in the hybrid RX correlation relate to the evolutionary phases (rise or decay) of the outburst, i.e. the $p\approx 1.3$ branch is achieved during the rise phase, while the $p\sim 0$ branch is established during the decay phase. However, this model is disfavored by observations in GRS 1739--278. In this source, both the rise and the decay phases are observed. But as shown in \autoref{fig:rxcorr2}, no any clear difference in the slope of the RX correlation among the two phases is observed. 

\citet*{Espinasse18} took another approach. They separated the radio-loud sources from the radio-quiet ones based on their RX correlations, and investigate the distribution of radio spectral indices within each subsample. Note that the radio-loud and -quiet ones correspond to respectively, the standard and the hybrid RX ones in our classification. \citet*{Espinasse18} found that the spectral index of the radio-quiet subsample (hybrid RX sources), $\alpha\approx -0.2$, is systematically lower than that of the radio-loud subsample (standard RX sources), $\alpha\approx 0.2$ (see also \citealt*{Brocksopp13} for a hint of such difference). This interpretation agrees with our data. According to their classification, GRS 1739--278 is a radio-quiet system, where a clustering of $\alpha$ at $\alpha\sim -0.2$ is observed in the hard state. We caution that epochs of even steeper radio spectrum (i.e. $\alpha\sim -1$) are observed in GRS 1739--278 in its hard state, although the spectral index of these observations suffers large uncertainties, see \autoref{fig:lc}.

The above two scenarios are motivated by observations. There is another one that is motivated by the progress on the fundamental properties of accretion physics, i.e. the radiative efficiency of hot accretion flow (\citealt*{Xie16}; see \citealt*{MM14, Cao14, Qiao15} for the $p\approx 1.3$ branch only, and \citealt*{Coriat11} for the efficiency requirement from observations). One advantage of this interpretation is that it is based on the truncated accretion--jet model \citep{Esin97, YN14}, which has been successfully applied to the hard state of BHBs. In this model, the synchrotron radio emission from a jet follows $\lr \propto \dot{M}_{\rm jet}^{\sim 1.4}$ \citep{Heinz03}, where $\dot{M}_{\rm jet}$ the mass loss rate into the jet. If the X-ray emission from hot accretion flow scales with mass accretion rate $\dot{M}$ as $\lx \propto \dot{M}^k$ (parameter $k$ characters the radiative efficiency in X-rays) and $\dot{M}_{\rm jet}\propto \dot{M}$, then we have $\lr \propto L_{\rm X}^{\sim 1.4/k}$ \citep{Heinz03, Coriat11}. In this picture, different slope in RX correlation is due to the difference in $k$, i.e. standard one has $k\approx 2.2$ \citep{Esin97, Merloni03}, flat $p\sim 0$ one has $k\gg 1$, and $p\approx 1.3$ has $k\approx 1$. Interestingly, such change in $k$ is indeed observed  in hot accretion flows, where depending on $\dot{M}$ three distinctive accretion modes are found \citep{Xie12}. We emphasis that the change in accretion mode will also result in a change in the spectral properties, which is confirmed in both BHBs and AGNs (see e.g., \citealt*{Yang15, Ruan19, Li19}).

This efficiency-related model by \citet*{Xie16} also predicts that the viscosity parameter of hot accretion flow, $\alpha_{\rm hot}$, should be small in hybrid RX sources (e.g., BHB H1743--322: \citealt*{Xie16}, and AGN NGC 7213: \citealt*{Xie16b}). Although $\alpha_{\rm hot}$ is difficult to measure, it can be crudely estimated from the critical luminosity of the hard-to-soft state transition $L_{\rm crit}$, since theoretically we have $\alpha_{\rm hot}\propto L_{\rm crit}^{\sim 1}$ \citep{Xie12, Xie16b}. Interestingly, both GRS 1739--278 and MAXI J1535--571 \citep{Russell19, Parikh19} are hybrid RX sources with state transitions observed in the mini-outbursts, thus they agree well with this predication.

\subsection{Disk--Jet Coupling During the Mini-Outbursts} \label{sec:miniburst}

So far the disk--jet coupling during the mini-outbursts has been investigated only in two BH transients: one is GRS 1739--278 in this work, the other is MAXI J1535--571 in \citet{Parikh19}. In both systems, the peak luminosities and durations of these mini-outbursts are at least one order of magnitude smaller than those of the main outburst, but still have state transitions at such low luminosities \citep{Yan17b, Parikh19}. Besides, the hard-to-soft state transition luminosity and the peak luminosity follows the same correlation that is established in the main outbursts of BHBs, implying that there is no intrinsic physical difference among these two types of outbursts \citep{Yan17b}.

The jet properties in those short-duration mini-outbursts are also similar to those in main outbursts. There is no radio observation of GRS 1739--278 during the hard state of the main outburst, but observations of MAXI J1535--571 \citep{Parikh19} suggest that both the main and the mini outbursts follow the hybrid RX correlation in the hard state. Moreover, considering the difference in the X-ray luminosity among main and mini outbursts, the $p\approx 1.3$ branch of the hybrid RX correlation may only exist in the main outburst, cf. the case of MAXI J1535--571 (\citealt*{Russell19}, admittedly there are only two data points).

\section*{Acknowledgements}
We appreciate the referee and Thomas Russell for careful reading and insightful suggestions. This work was supported in part by the National Program on Key R\&D Project of China (Grants 2016YFA0400804), the Natural Science Foundation of China (grants 11763002, 11773055, 11873074, U1838203, U1938114 and U1931203), the K.C. Wong Education Foundation of CAS, and the Key Research Program of Frontier Sciences of CAS (No. QYZDJ-SSW-SYS008). F.G.X. and Z.Y. were also supported in part by the Youth Innovation Promotion Association of CAS (ids. 2016243 and 2020000). 

\vspace{5mm}
\facilities{JVLA, Swift}
\software{XSPEC, CASA, Difmap}

\end{document}